\documentclass[epj,nopacs]{svjour}
\usepackage{graphicx}
\usepackage{array}
\usepackage{amsmath,amssymb}
\newcommand{\kf}{k_{\scriptscriptstyle\rm F}}

\newcommand{\beqn}{\begin{equation}}
\newcommand{\eeqn}{\end{equation}}
\newcommand{\bea}{\begin{eqnarray}}
\newcommand{\eea}{\end{eqnarray}}
\newcommand{\ba}{\begin{align}}
\newcommand{\ea}{\end{align}}

\newcommand{\ie}{\textit{i.e.}}
\newcommand{\eg}{\textit{e.g.}}
\newcommand{\cf}{\textit{cf.}}

\newcommand{\I}{\item}

\newcommand{\bi}{\begin{itemize}}
\newcommand{\ei}{\end{itemize}}
\newcommand{\be}{\begin{enumerate}}
\newcommand{\ee}{\end{enumerate}}

\newcommand{\abinitio}{\textit{ab initio}}
\newcommand{\xvec}{\textbf{x}}
\newcommand{\Vext}{v_{\rm ext}}
\newcommand{\bfx}{{\bf x}}

\newcommand{\wh}{\widehat}
\newcommand{\Hhat}{\widehat H}
\newcommand{\Qhat}{\widehat Q}

\newcommand{\la}{\langle}
\newcommand{\ra}{\rangle}

\newcommand{\HBRSThat}{\widehat H_{\mathrm{BRST}}}

\begin{document}
\title{Turning the nuclear energy density functional method into\\ a proper effective field theory: reflections}
\author{R. J. Furnstahl\inst{1} %\and Second author\inst{2}% etc
}                     % Do not remove
\institute{Department of Physics, Ohio State University, Columbus, OH 43210, USA}
\date{Received: date / Revised version: date}
\abstract{
Nuclear energy density functionals (EDFs) have a long history of success in reproducing
properties of nuclei across the table of the nuclides.
They capture quantitatively the emergent features of bound nuclei, such as nuclear saturation
and pairing, yet greater accuracy and improved uncertainty quantification are actively sought.
Implementations of phenomenological EDFs are suggestive of effective field theory (EFT)
formulations and there are hints of an underlying power counting.
Multiple paths are possible in trying to turn the nuclear EDF method into a proper EFT.
I comment on the current situation and speculate 
on how to proceed using an effective action formulation.
\PACS{
      {PACS-key}{describing text of that key}   \and
      {PACS-key}{describing text of that key}
     } % end of PACS codes
} %end of abstract
\maketitle
%

%%%%%%%%%%%%%%%%%%%%%%%%%%%%%%%%%%%%%%%%%%%%%%%%%%%%%%%%%%%%%%%%%%%%%%%%%%%%%%%%%%%%%
%%%%%%%%%%%%%%%%%%%%%%%%%%%%%%%%%%%%%%%%%%%%%%%%%%%%%%%%%%%%%%%%%%%%%%%%%%%%%%%%%%%%%

\section{Introduction}
\label{sec:intro}

The many levels of emergent phenomena in strong interaction physics have led to a corresponding 
set of models that have successfully described experiment at each energy scale 
(see Table~\ref{tab:emergent}).  
The tension between simplified models for these phenomena and a reductive description via
an underlying theory has been bridged in almost all cases by an effective field theory (EFT).%
\footnote{The term ``effective field theory'' is often restricted to mean a local Lagrangian
formulation of a low-energy theory, with ``effective theory'' a more general designation.
We mostly have in mind strict EFTs but for convenience will use that designation even for the more general cases.}
An EFT is characterized by a selection of the relevant degrees of freedom, an identification of
symmetries to enforce,
and a power counting that establishes an order-by-order expansion, incorporating fine-tuning 
as needed~\cite{Georgi:1994qn,Burgess:2007pt,Furnstahl:2008df}.
One might observe that behind every successful emergent phenomenology there is an EFT 
(or possibly more than one) waiting to be uncovered.
The insight, inspiration, and validation from a more microscopic description 
can be absorbed by the EFT
or guide its construction through matching.  In turn, the EFT provides clarity to models, computational
simplification, and an opportunity for greater precision.
     
The one clear gap in the EFT hierarchy of Table~\ref{tab:emergent} is for describing 
ground state properties of nuclei.
Energy density functionals (EDFs) remain the phenomenological method of choice for these 
observables across the bulk of the table of nuclides~\cite{Bender:2003jk,10.1088/2053-2563/aae0ed}.
This is despite great progress in the reach of ab initio methods using chiral EFT Hamiltonians, for there
is still much of the table uncovered and the accuracy for the ground states is not comparable although
generally superior for low-lying spectroscopy.
The form and organization of these EDFs are suggestive of EFTs, but there remain major holes
to fill and questions to answer before a robust theory is available.
My goal here is to survey where we stand and the possibilities for filling the holes.
A theme I will adopt is that we should use the phenomenological success as a close, 
though not infallible, guide to formulating an appropriate EFT.
I will touch upon many promising avenues being pursued and then outline 
a particular path that has not been explored in depth.
I caution the reader that this is a selection and assessment based on personal prejudices, 
not a comprehensive review, and therefore some approaches will not be addressed and 
references to the literature will not be exhaustive.

\begin{table*}[th!]
  \begin{center}
    \caption{Emergent phenomenologies and the EFTs that have supplanted them. }
    \label{tab:emergent}
    \begin{tabular}{c | c | p{6cm}}
    \hline\noalign{\smallskip}
     Emergent phenomena &   Phenomenology  &   
     \multicolumn{1}{c}{Effective field theory (EFT)} \\ \hline\hline 
    nucleons as confined quarks/gluons &    constituent quark model &  chiral quark model~\cite{Manohar:1983md} \\ \hline
   inter-nucleon forces  &   meson exchange models 
       & chiral EFT with nucleons, [$\Delta$’s,] pions~\cite{Epelbaum:2008ga} \\ \hline
  \multicolumn{1}{>{\centering}p{5cm}|}{universal large-scattering-length physics and halo nuclei \hfill} & 
  \multicolumn{1}{>{\centering}p{5cm}|}{effective range expansion and   cluster models}  &  
  pionless EFT: nucleons only~\cite{Epelbaum:2008ga} or
                            nucleons and clusters (halo)~\cite{Hammer:2017tjm} \\ \hline
    saturation, pairing, shell structure &  energy density functionals    & \textit{What goes here???}      \\ \hline
    low-lying excitations, superfluidity   &  nuclei as Fermi liquids~\cite{migdal1967theory}
    &  EFT at the Fermi surface~\cite{Polchinski:1992ed,Shankar:1993pf} \\ \hline
    \multicolumn{1}{>{\centering}p{5cm}|}{collective rotational/vibrational motions of deformed nuclei}  
    &  collective models &  effective theory for deformed nuclei: systematic collective dofs~\cite{Papenbrock:2013cra} \\ \hline
    \end{tabular}
  \end{center}
\end{table*}

%%%%%%%%%%%%%%%%%%%%%%%%%%%%%%%%%%%%%%%%%%%%%%%%%%%%%%%%%%%%%%%%%%%%%%%%%%%%%%%%%%%%%

\subsection{Emergent features of nuclei captured by EDFs}

As I intend to use the phenomenological successes of EDFs as a guide, let me briefly
review some of the emergent features of nuclei manifested by empirical EDFs such as Skyrme models
or Gogny models~\cite{Bender:2003jk,10.1088/2053-2563/aae0ed}.  
Nuclear saturation is characterized by quantitatively precise liquid drop systematics 
overlaid with a regular shell structure~\cite{Bulgac:2017bho}.
Every EDF that is successful in fitting nuclear masses and radii predicts almost the same 
nuclear matter saturation properties: very close binding energies and equilibrium densities, with greater
spread in the symmetry energy and compressibility.  
(There are some systematic differences between nonrelativistic and
covariant functionals, possibly because of the incompleteness of their respective functional forms,
but these differences are small.)
Bulk deformations are well described as are separation energies.

To describe pairing, 
modern EDFs implement some variety of the Har\-tree-Fock-Bogo\-liubov (HFB) formalism.
Pairing is manifested  in the form of an even-odd staggering of masses, with the magnitude
reflecting the size of the pairing gap. 
The specification of the force or functional in the pairing channel is usually much less detailed
than in the particle-hole channel.
EDFs that are not derived from a Hamiltonian generally treat these channels independently while others
such as Gogny implementations originate from an effective force that may be the same in both channels.
We will return to the apparent lower resolution in the pairing channel
later, as it is a relevant element to consider for a EFT implementation.
Finally, mean collective properties of excitations such as giant resonances are generally well described in a
small-amplitude time-dependent mean-field approach (RPA or QRPA), 
although not damping or fragmentation~\cite{harakeh2001giant}.
Note that these features do not exhaust the scope of EDFs (see ref.~\cite{10.1088/2053-2563/aae0ed}
for a recent overview), but are sufficient for our discussion.

The tension between underlying complexity and emergent simplicity is evident when comparing
the functionals to microscopic inter-nucleon interactions based on chiral EFT.
On the one hand, there is evidence that the naturalness of low-energy constants in the
chiral EFT Lagrangians, when scaled according to naive dimensional analysis, 
is inherited by parameter values in Skyrme and covariant functionals~\cite{Friar:1995dt,Rusnak:1997dj,Kortelainen:2010dt}
(with the caveat that while the evidence is compelling it is not absolutely convincing because of the limited orders available).
Thus the EDFs in this way appear to reflect underlying chiral physics.
At the same time, there is no overt support in the functionals for chiral symmetry constraints~\cite{Furnstahl:2010wc}.
Furthermore, multiple studies over the years have shown that relatively few
parameter combinations determine most of the physics~\cite{Vautherin:1971aw,Furnstahl:1999rm,Bertsch:2004us,Bulgac:2017bho}, for example, from doing a singular
value decomposition to uncover the dominant combinations.
These combinations
reflect emergent saturation properties, not the chiral Lagrangian parameters (which are themselves
interpretable
via resonance saturation in terms of meson exchange~\cite{Epelbaum:2001fm}).
The limited role of pions in mean-field descriptions is perhaps clearest in covariant functionals,
where they appear in a defining Lagrangian, 
but are suppressed because of spin and isospin averaging in the bulk.
An analogous averaging argument for the dominance of $SU(4)$-invariant interactions was made recently by Lu et al.\ while considering ``essential elements for nuclear binding'': these elements did not include pions~\cite{Lu:2018bat}.
Are we seeing phenomenological signals that pionic degrees of freedom are not optimal 
(or necessary) for nuclear EDFs?

One might question whether the nuclear properties I have identified should be called ``emergent''
based on how that term is used elsewhere.
For my purposes it means that the phenomena reflect a complexity or collectivity not evident in the
degrees of freedom (dofs) of the underlying Hamiltonian, such that a more transparent 
description would require different dofs. 
Nuclear saturation and pairing are certainly emergent from the viewpoint of QCD, 
where even the nucleons themselves are highly nontrivial consequences of the QCD Hamiltonian.
But with respect to chiral EFT they are not obviously emergent by my definition. 
Rather, one might say that it is evident from the mid-range attraction and short-range repulsion built
into fitting phase shifts, plus a repulsive three-body force, 
that there will be saturation (\ie, a liquid phase); that low-lying collective
modes are inevitable when thinking of vibrating liquid drops; and that attraction near the Fermi surface
always means there will be pairing. 
While this may all be true, the \emph{quantitative} properties are very finely determined.
We do know that
if we fit chiral EFT parameters to few-body properties, 
we \emph{will} get saturation, just not necessarily
in the right place (\eg, see~\cite{Drischler:2017wtt}).
And it is a precise determination that is key to a quantitative, not just qualitative,
description of nuclear properties.

%%%%%%%%%%%%%%%%%%%%%%%%%%%%%%%%%%%%%%%%%%%%%%%%%%%%%%%%%%%%%%%%%%%%%%%%%%%%%%%%%%%%%

\subsection{Why should we try to do better?}  \label{subsec:better}

The motivations for doing better than the current empirical EDFs are well
aligned with what an EFT could (in principle) do for us.  
\begin{itemize}
  \item The breakdown and failure mode for EDFs is unclear.  For example, \emph{should} they
  work all the way
  to the drip\-lines as currently applied 
  or should they degrade when bulk systematics become less dominant?  
  The internalization
  of the breakdown scale is a hallmark of the EFT approach.

  \item 
  How do we improve EDFs in a controlled manner?
  Are density dependencies too simplistic? How do you know?
  How should we organize possible terms in the EDF?
  All of these related questions are addressed by EFT power counting.

  \item More accuracy is wanted, \eg, for r-process calculations; is this even possible?  That is,
  what is the theoretical limit of accuracy?  This is again an EFT feature because there
  is an expansion parameter that enables a quantification of missing physics~\cite{Melendez:2019izc}.

  \item There is apparent model dependence seen upon comparing different parameterizations of
  empirical EDFs.  How should this
  be interpreted?  A consequence is that extrapolations to the driplines, 
  to large $A$, and to high density are at least to some extent uncontrolled.
  The completeness of an operator basis in an EFT addresses issues of model dependence.

  \item An important contemporary theme in nuclear theory is the robust estimation of theoretical 
  uncertainties, which ties in with some of the other issues.  
  There have been significant advances in EDF uncertainty quantification (UQ)~\cite{10.1088/2053-2563/aae0ed}, 
  but a proper EFT would add important prior knowledge from the expansion convergence pattern~\cite{Melendez:2019izc}.

  \item What observables can be calculated in a controlled way and how does one know how
  to couple to external currents (\eg, for electroweak properties)?  
  Within formal density functional theory (DFT),%
  \footnote{The important distinction between DFT as formalized for the Coulomb many-body problem and the
  nuclear EDF approach has been stressed by Duguet and collaborators~\cite{Duguet:2010cv}. We will not
  address this issue explictly until we consider zero modes in sect.~\ref{subsec:zeromodes}.
  Until then we will generally use DFT and EDF interchangeably.} 
  certain quantities such as single-particle levels
  are not guaranteed to accurately correspond to physical quantities, although they are still extracted
  from EDFs. 
  The consistent construction of a Hamiltonian and currents is
  another fundamental feature of EFTs.
  The formulation of the EDF as an EFT (such as in an effective action framework, see below) leads to a
  clear identification of accessible observables and a guide to extensions for other
  observables (\eg, for single-particle levels
  one can relate the full and Kohn-Sham Green's functions~\cite{Bhattacharyya:2004aw}).

  \item Finally, we desire connections to other nuclear EFTs in the hierarchy 
  leading to quantum chromodynamics (QCD).
  The search for intersections raises many questions we will be able to address with an EFT:
   When is pion physics resolved?  Does near-unitarity of nuclear forces matter?  What is the connection
   to the many-body forces that play a vital role in nuclear saturation in chiral EFT?
\end{itemize}
Together these form a strong case for seeking a proper EFT formulation of EDFs.
So having decided an EFT is worth pursuing, how should we proceed?
We could try top-down, starting from ab initio methods, or bottom-up with a ``general''
functional, matching to experiment and/or ab initio calculations.
A hybrid plan on the way to a proper EFT 
could be to extend or modify existing EDF forms in (semi-)controlled way, 
using the microscopic many-body theory for guidance.
There has been progress in nuclear many-body theory that touches all of these paths.
I will first comment on implications of this progress before turning to my current
favorite bottom-up plan.

%%%%%%%%%%%%%%%%%%%%%%%%%%%%%%%%%%%%%%%%%%%%%%%%%%%%%%%%%%%%%%%%%%%%%%%%%%%%%%%%%%%%%
%%%%%%%%%%%%%%%%%%%%%%%%%%%%%%%%%%%%%%%%%%%%%%%%%%%%%%%%%%%%%%%%%%%%%%%%%%%%%%%%%%%%%

\section{Progress report}  \label{sec:progress}

About a decade ago I co-authored a summary and for\-ward-looking review
entitled ``Toward ab initio density functional theory for nuclei''~\cite{Drut:2009ce}.
The focus was how to go from ab initio microscopic methods to a DFT,
with some discussion of EFT for EDFs.
The article described two different general approaches: through many-body perturbation theory (MBPT),
possibly resummed, or through an effective action formalism.  
Much in the nuclear theory landscape has changed since then.  
Here we will consider elements of the contemporary landscape in light of the goal of a proper
EFT for EDFs.

%%%%%%%%%%%%%%%%%%%%%%%%%%%%%%%%%%%%%%%%%%%%%%%%%%%%%%%%%%%%%%%%%%%%%%%%%%%%%%%%%%%%%

\subsection{Lattice quantum chromodynamics (LQCD)}  \label{subsec:LQCD}

Numerical calculation on a space-time lattice is a well-established method for 
solving QCD in the confinement regime (functional renormalization group methods
also have many prom\-ising aspects and Schwinger-Dyson methods offer good 
QCD-motivated phenomenology).
LQCD is now capable of first-principles calculations of hadron masses and decay constants, with
controlled errors at the percent level.
Despite this progress, not long ago it was thought that 
accurately calculating the residual force between nucleons was beyond reach in principle
because of signal-to-noise arguments.
However, the worst of those problems have been overcome
and there are active calculations of multi-hadron systems from multiple groups~\cite{Davoudi:2019jjk}.

There are still formidable technical problems and qualitative disagreements to work out for
the two major methods being employed (\eg, see~\cite{Iritani:2017rlk,Beane:2017edf,Iritani:2018vfn}).  
But even if resolved, at this stage it appears the most likely scenario for the role
of LQCD will be to match to chiral EFT
to determine low-energy constants and calibrate experimentally challenging 
phenomena such as three-neutron forces and strangeness,
rather than providing direct input to an EDF formulation.
This is consistent with the general paradigm based on the renormalization group (RG)
that a tower of EFTs replaces
direct calculations with the underlying theory.
The more reductive frameworks are relevant for proofs-of-principle and to 
identify missing physics, but for tractable precision
calculations and cleaner physical insight, an EFT at the appropriate resolution
for nuclei near their ground states is the way to go.

So what is the appropriate resolution?
Given that the Fermi momentum inside of a nucleus is larger than the pion mass, one
would naively imagine that the EFT must include the pion as a resolved degree of freedom,
which implies chiral EFT, which we consider next.

%%%%%%%%%%%%%%%%%%%%%%%%%%%%%%%%%%%%%%%%%%%%%%%%%%%%%%%%%%%%%%%%%%%%%%%%%%%%%%%%%%%%%

\subsection{Ab initio with chiral EFT} \label{subsec:abinitio}

We have entered an era of precision calculations of nuclear structure and
reactions, which seek to address the full table of nuclides and astrophysical
systems such as neutron stars.  
The precision era 
has been enabled by advances in theoretical methods,
both conceptual and algorithmic; computational capabilities; and
enhanced confrontation with experiment. 
How should we exploit these new capabilities in the context of improved EDFs?
One possible route is to use direct extensions
  of ab initio many-body methods in a top-down manner via MBPT~\cite{Drut:2009ce}. 
Ab initio in this context means calculations of nuclei based on a free-space Hamiltonian with  
proton and neutron degrees of freedom that is fit to (at least) two-body and few-body data
(usually scattering phase shifts for the former and bound-state properties for the latter).

This route is conceivable because
the last decade has seen astounding progress in the reach of ab initio methods. 
Within the last half-dozen years, the upper bound for 5\% calculations with realistic Ham\-iltonians (which generally require three-body forces) has grown from $A=14$ to $A$ of order 80, with calculations
in selected regions reaching even higher~\cite{Hergert:2018wmx}.
The Hamiltonians in these calculations 
are generally from chiral EFT with the pionic interactions converted to potentials,
for example by decoupling the pion sector via unitary transformations~\cite{Epelbaum:2008ga,Machleidt:2011zz}. 
Although long recognized as important, $\Delta$ degrees of freedom are just now being included in
mainstream potentials.

A wide range of complementary computational methods are used to implement chiral EFT for nuclei~\cite{Hjorth-Jensen:2017gss}. 
Each exhibits characteristic strengths.
Lattice EFT, in which spacetime is discretized as in LQCD and pion interactions are replaced by
auxiliary fields, is naturally suited to described clustering.
Stochastically improved wave functions with the auxiliary field diffusion Monte Carlo method
can handle short-range correlations that are problematic in general.
The no-core shell model (NCSM) uses large-scale diagonalization, which gives complete excitation spectra
that demonstrate the emergence of collective rotational bands~\cite{Maris:2014jha}.
(There have also been many developments of both shell model EFT and EFT in an oscillator basis.)
Coupled cluster methods and the in-medium renormalization group,
which sum infinite-order sets of diagrams through nonlinear equations and flow equations, respectively,
have favorable scaling to large nuclei and effective interactions for
the phenomenological shell model emerge naturally.  
Self-consistent Green's function (SCGF) methods enable conserving approximations and direct access to
nuclear spectral functions.

The proliferation of these methods has been paralleled by a recent proliferation of chiral
EFT Hamiltonians.
These Hamiltonians have one of two classes of physics content, with or without $\Delta$s,
but otherwise only differ in the type of regularization, which is used to cut off high momenta.
In principle differences in the regularization should be absorbed in the process of renormalization,
but the current implementations are not strictly renormalizable in the sense of being cutoff
independent at a fixed order.  An immediate consequence is that there are regulator artifacts that
can degrade the EFT performance.  More generally problematic is that the renormalization group (RG)
cannot be directly applied to these interactions within the EFT.
It is also important to note that the power counting for these potentials, which determines what contributes at
each EFT order, is done in free space and then the many-body Schr\"odinger equation is solved
as exactly as possible using the resulting potential.
So there is no accounting for possible altered power counting at sufficiently high density.

The paradigm for descending a tower of EFTs is to match an emergent EFT 
to an underlying one at a scale where
degrees of freedom are eliminated.  This is where naive dimensional analysis estimates of low-energy
constants are expected to be good.
Then one uses the renormalization group to evolve to an appropriate lower scale for phenomena
under consideration.
For free-space nuclear Hamiltonians, RG methods such as the similarity RG are used to integrate out
high-momentum modes, which makes the interaction softer, meaning more perturbative.
In general, such a softening should be 
associated with having a more appropriate resolution for the problem at hand
because the coupling between states that appears in perturbative expansions is reduced.
The practical counterbalance in the nuclear case is the growth of many-body forces, which are difficult
to treat with ab initio methods; this has limited the use of the SRG and similar approaches
to only moderate softening.

In all, the prospects are good for ab initio methods improving their coverage of the table of
nuclides, with the advantage of consistent currents.
However, it is uncertain at best whether they will be able to challenge the accuracy of the
phenomenological EDFs for masses.  
Calculations at the percent level are feasible, but for $^{208}$Pb, for example,
that would mean an error of order 15 MeV in the binding energy, while the best EDFs are at the
1~MeV level or below.  This may be an insurmountable barrier, but time will tell.  
It is noteworthy that the best reproduction
of nuclear masses and radii is achieved when the chiral EFT low-energy constants are determined by
fits that include nuclei well above $A=4$.  
In essence, they must be fine-tuned to saturation properties, as is done with EDFs.

An alternative to directly solving for nuclear properties using chiral EFT and ab initio methods is to use them to \emph{inform}
nuclear EDFs (see also sect.~\ref{subsec:no_pions}). 
Duguet and collaborators~\cite{Duguet:2014jja,Duguet:2015yle,Duguet:2015nna,Ripoche:2016afk}
have developed a strategy to use MBPT and coupled cluster theory as frameworks for constructing novel 
parameterizations of nuclear EDFs that overcome problematic issues for EDFs.
A key element is a proper treatment of symmetry breaking and restoration (see sect.~\ref{subsec:zeromodes}).
The goal is ``to guide the construction of safe, explicitly correlated and systematically improvable parameterizations.''
We will return to the critical issue of broken symmetries later.

%%%%%%%%%%%%%%%%%%%%%%%%%%%%%%%%%%%%%%%%%%%%%%%%%%%%%%%%%%%%%%%%%%%%%%%%%%%%%%%%%%%%%

\subsection{Using phase space estimates for power counting}

For any quantum field theory, a
regularization scheme and scale must be introduced to define the
theory. To regulate divergent loops, chiral EFT introduces
cutoff regulators, which become an intrinsic part of the nuclear potentials.
With the Weinberg power counting scheme, which is used in essentially all
microscopic calculations to date,  iteration of the potential beyond
leading order generates divergences without all the
corresponding counterterms. As a result, the EFT is no longer strictly
renormalizable and, as already noted, there are significant regulator artifacts.
Until recently, almost all calculations used potentials with similar
non-local regulators.

However, there are now local regulators for use with quantum Monte
Carlo calculations~\cite{Gezerlis:2013ipa,Piarulli:2014bda} and semi-local regulators
designed for improved power-counting behavior in free space~\cite{Epelbaum:2014efa,Reinert:2017usi}.
Recent work has studied 
the impact of different regulators for finite density calculations~\cite{Dyhdalo:2016ygz}.
The strategy was to use uniform matter as a testbed and work 
in MBPT~\cite{Hebeler:2010xb}.
A Monte-Carlo sampling method was developed to probe the interaction phase
space in energy integrals using a variety of regulators,
which revealed some striking differences between different regulators.
(The observed impact of the regulators led to revisiting the implementation of the density
matrix expansion, which had stalled in previous attempts that did not include regulators;
see the next section.)
A subsequent phase-space-based effort with the same sampling approach was aimed at 
developing and validating a many-body power counting for softened potentials
in uniform matter~\cite{Dyhdalo:2017gyl}.
This power counting completely alters the conventional hole-line-expansion 
power counting developed for hard potentials.
At densities near nuclear matter saturation, Pauli blocking and decoupling of
high-momentum modes combine to dramatically
change the counting one expects in free-space (\eg, fine-tuning
in the S waves goes away).

The diagnostic sampling tools to analyze many-body contributions
showed that quantitative estimates are feasible in the particle-particle
channel~\cite{Dyhdalo:2017gyl}.
This approach is being extended to the two-body particle-hole
channels, where a power counting prescription is less clear, and to more completely
examine the role of three- and higher-body forces.
Such empirical studies using the sampling technique, 
if applied to the effective action formalism advocated
in sect.~\ref{sec:favorite}, can give us direct insight and validation into how 
power counting should work for
a nuclear EDF.

%%%%%%%%%%%%%%%%%%%%%%%%%%%%%%%%%%%%%%%%%%%%%%%%%%%%%%%%%%%%%%%%%%%%%%%%%%%%%%%%%%%%%

\subsection{Generating an EDF with pions using the DME} \label{subsec:hybrid}

The perturbative in-medium results from low-mo\-men\-tum (\eg, RG-evolved) potentials could be taken
to suggest that pion as dofs are still appropriate at nuclear matter densities but that
an alternative EFT power counting from chiral EFT is needed.
Kaiser and collaborators
have proposed a \emph{perturbative} chiral theory approach to nuclear matter
and then to finite nuclei through an EDF functional~\cite{Kaiser:2009me,Holt:2013fwa}.
They consider Lagrangians both for nucleons and pions and for nucleons, pions, and $\Delta$'s, and fit 
parameters to nuclear saturation properties.  
They construct a loop expansion for the nuclear matter energy
per particle, which leads to an energy expansion of the form
\begin{align}
  E(\kf) &= \sum_{n=2}^\infty \kf^n \, f_n(\kf/m_\pi,\Delta/m_\pi) \;, \\
   \Delta &\equiv M_\Delta - M_N \approx 300\,\mbox{MeV} \;,
\end{align}
where each $f_n$ is determined from a finite number of
in-medium Feynman diagrams, which incorporate the long-distance
physics.
All powers of $\kf/m_\pi$ and $\Delta/m_\pi$ are kept in the $f_n$'s
because these ratios are not small quantities~\cite{Holt:2013fwa}.

A semi-quantitative description of nuclear matter is
found even with just the lowest two terms without 
$\Delta$'s and adding $\Delta$'s brings uniform improvement
(\eg, in the neutron matter equation of state).
By applying the density matrix expansion (DME) in momentum
space to this expansion they 
derive a Skyrme-like EDF
for nuclei~\cite{Holt:2013fwa}.
The DME provides a general way to map nonlocal functionals into local ones 
by converting one-body density matrices (OBDMs) into local densities.  
In particular, the nonlocality in the OBDMs arising from finite-range potentials
is factorized into products of local densities multiplied by density-dependent couplings. 
The resulting EDF has qualitatively correct features but the
accuracy is not competitive to phenomenological EDFs.

A more phenomenological approach is to apply 
the DME to long-range pion terms following the Weinberg
power-counting expansion for the potential, then to merge the resulting functional with a 
conventional Skyrme functional and re-optimize.
The idea is to inject novel density dependence from the long-range pion physics
while adjusting the Skyrme parameters to absorb short-distance physics.
Table~\ref{table:masses} (adapted from ref.~\cite{NavarroPerez:2018tme}) shows intriguing results from
the application of the coordinate-space, regulated DME formulation by Dyhdalo et al.~\cite{Dyhdalo:2016tgx}
to NN and NNN diagrams.
In particular, with the same fitting protocol and the same free parameters, the r.m.s.\ residual
for binding energies improved significantly compared to the UNEDF-2 reference when 
chiral NN contributions
are added, particularly with an explicit $\Delta$ (\eg, to 1.41\,MeV for NLO$\Delta$ and 
1.26\,MeV for N2LO$\Delta$ compared to 1.98\,MeV for UNEDF-2).  
But after including three-body contributions, the result does not improve or even
becomes significantly worse (compare N2LO$\Delta$+3N to N2LO$\Delta$).
Whether the superior performance of the functional was due to including correct density dependence
from pion exchange or to a better optimization from the new terms is still under
investigation.

\begin{figure}[t!]
        \begin{center}
        \includegraphics[width=0.98\columnwidth]{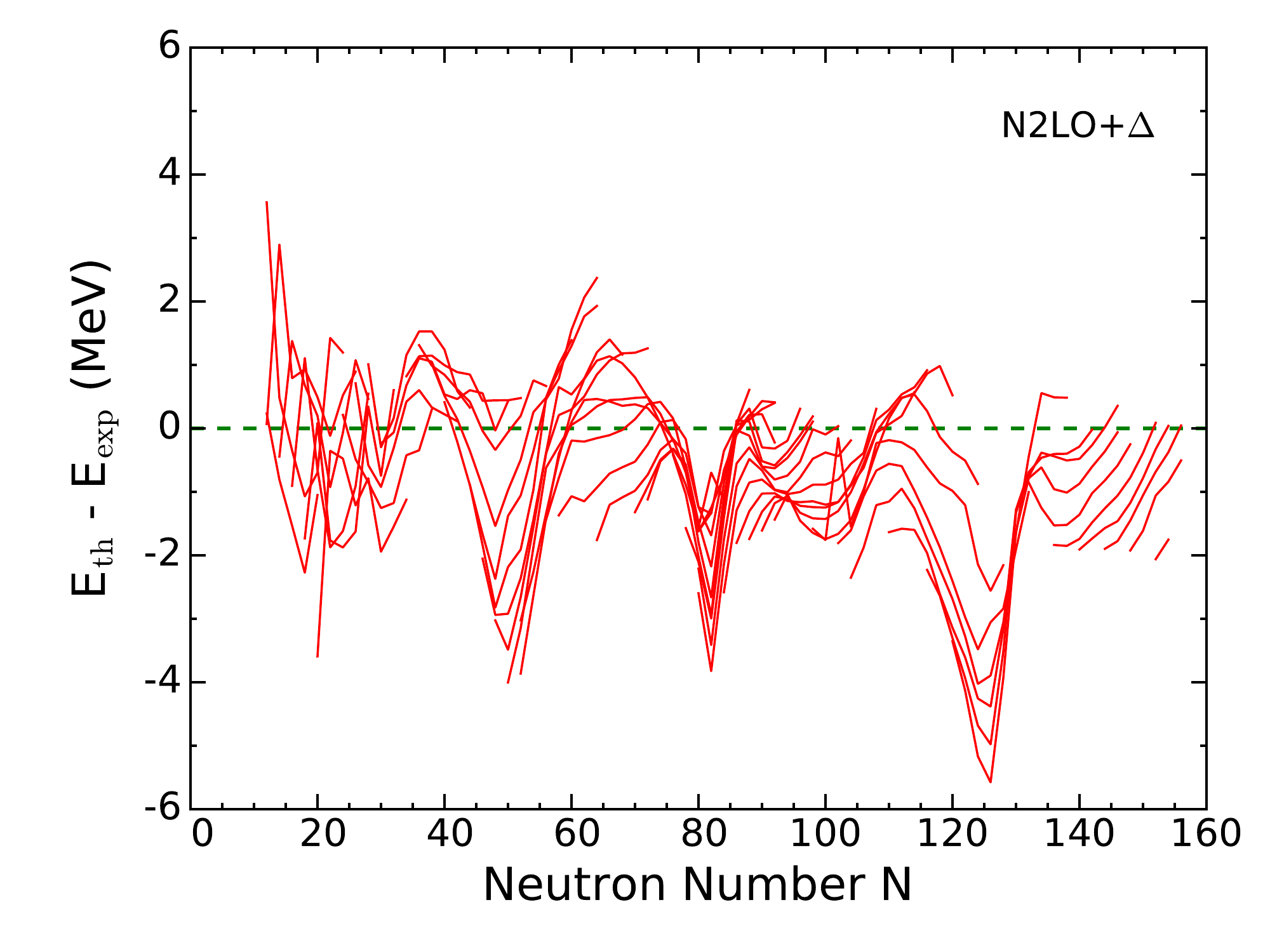}
        \caption{Residuals for N2LO$\Delta$ (see Table~\ref{table:masses})~\cite{NavarroPerez:2018tme}.}
        \label{fig:masses_N2LOD}
        \end{center}
\end{figure}

The plot of binding energy residuals in fig.~\ref{fig:masses_N2LOD} for the most favorable case from
ref.~\cite{NavarroPerez:2018tme} shows definite improvement when compared to the UNEDF-2 standard,
but also manifests similar systematic deviations from zero.  
At least some of these deviations are attributable to ``beyond mean field'' physics, meaning that one needs to go beyond 
the standard Hartree-Fock-Bogoliubov (HFB) energy density functional (EDF), such as doing
symmetry restoration and incorporating coupling to vibrational modes.
An example of how such corrections reduce the pattern of residuals seen in fig.~\ref{fig:masses_N2LOD}
is given in ref.~\cite{Goriely:2016uhb} (see also~\cite{Bender:2004tg}).
The bottom line is that we have clear evidence that the performance of the standard Skyrme functional
can be improved, but whether explicit pions are needed is still open.
An interesting possibility is that the long-range pion physics is needed at the mean-field 
(Hartree-Fock or even just Hartree) level, but not at higher orders in MBPT (\eg, see ref.~\cite{Zhang:2018tqf}). 

\begin{table}[t!]
      \caption{Root mean square (r.m.s.) deviations between experimental and theoretical binding 
        energies in MeV~\cite{NavarroPerez:2018tme}. 
        The ordering is according to free-space Weinberg power counting, with and without three-body
        forces ($3N$) and with and without $\Delta$s.
        }
      \label{table:masses}
      \begin{center}
      \begin{small}
      \begin{tabular}{cc}
            EDF             & r.m.s.\ residual (MeV)  \\
            \hline                
       UNEDF-2                 & 1.98  \\
         LO                 & 1.99  \\
        NLO                  & 2.02  \\
       N2LO                  & 1.57  \\
      N2LO+3N          & 1.58  \\
       NLO$\Delta$          & 1.41  \\
       NLO$\Delta$+ 3N & 1.46  \\
       N2LO$\Delta$          & 1.26  \\
       N2LO$\Delta$ + 3N & 1.72  \\
      \end{tabular}
      \end{small}
      \end{center}
\end{table}

%%%%%%%%%%%%%%%%%%%%%%%%%%%%%%%%%%%%%%%%%%%%%%%%%%%%%%%%%%%%%%%%%%%%%%%%%%%%%%%%%%%%%

\subsection{EFT and EDF without explicit pions} \label{subsec:no_pions}

Multiple paths are being followed to connect EFT and EDFs that do not invoke pions
and chiral symmetry.
An EFT with free-space nucleon fields only, known generically as pionless EFT,
is a much better understood and controlled framework than chiral EFT, which in its current
implementation is faced with intertwined open issues and controversies of power counting,
renormalizability, perturbative versus nonperturbative resummations, and
regulator dependence.
Past explorations of an EFT expansion for nuclear energy functionals by the author and collaborators 
were based on a controlled low-density expansion in pionless EFT, whereby a free-space momentum
expansion translates into an expansion in the Fermi momentum $\kf$ divided by a short-distance 
scale (such as the inverse range of the potential)~\cite{Hammer:2000xg,Furnstahl:2000we,Furnstahl:2001xq,Furnstahl:2002gt,Puglia:2002vk,Bhattacharyya:2004qm,Bhattacharyya:2004aw,Furnstahl:2006pa,Drut:2009ce}.
The form of the functional in the local density approximation has terms with
powers of the density just like the Skyrme-type functionals but with additional terms at
higher orders.
The effective action for local, composite operators can be used to connect to DFT, as 
I will revisit in  sect.~\ref{sec:favorite}.

Subsequently there have been interesting recent developments by others based on pionless EFT
but going beyond the low-density perturbative expansion and by connecting to EDFs beyond the
mean field level.
These include constraining terms in the functional by requiring renormalizability~\cite{Moghrabi:2013fwa,Yang:2017shi}
or going beyond a perturbative expansion by considering approximate self-energies~\cite{Boulet:2019wfd}.
A particularly compelling idea is to push the expansion of nuclei about the unitarity limit that
dictates the universal physics manifested in cold atoms and dilute neutron gases~\cite{Lacroix:2016dfs,Konig:2016utl,vanKolck:2018vzl}.
See refs.~\cite{Grasso:2016gls,Grasso:2018pen} for a review of these ideas. 
At this stage these are all worthwhile avenues to pursue; none of them is conclusive as yet.

%%%%%%%%%%%%%%%%%%%%%%%%%%%%%%%%%%%%%%%%%%%%%%%%%%%%%%%%%%%%%%%%%%%%%%%%%%%%%%%%%%%%%

There are several related directions motivated by EFT principles being pursued by
Dobaczewski and collaborators.  The first was to extend existing Skyrme functionals with additional
terms in a controlled, order-by-order gradient expansion~\cite{Carlsson:2008gm}.
However this does not overcome the evident limitations of the Skyrme functional form.
Next there is the formulation in terms of pseudopotentials, 
which specify the EDF upon folding with an uncorrelated Slater determinant,
which is found self-consistently.  Thus the full functional is given within the Hartree-Fock (or HFB)
approximation.
The secret of the EFT approach based on local Lagrangians is due to the expectation (not
generally proved) that the most general Lagrangian consistent with the symmetries of the system
leads to the most general S-matrix.  Thus a complete operator basis in the Lagrangian ensures
a model-independent formulation because everything possible can be accommodated.
The question is: can the pseudopotential formulation provide the analogous general 
blueprint for a nuclear energy functional?  If so, we would have a framework for a proper EFT for EDFs. 

In refs.~\cite{Raimondi:2011pz,Dobaczewski:2012cv,Raimondi:2014iia}, pseudopotentials, both zero-range and finite-range, are derived including all possible terms in a derivative expansion, following a well-defined
power counting that uses an identified in-medium scale.
An interesting argument is made that the resolution scale of chiral EFT is much
higher than is needed.  In particular, rather than using momentum $k \lesssim 2m_\pi$ or $\kf$,
one should consider $\delta k$ to dissociate a nucleon:
\beqn
    \delta E_{\textrm{kin}} = \frac{\hbar^2\kf \delta k}{M}
      \approx 0.25\hbar c\, \delta k \approx 8\,\mbox{MeV}
      \;,
\eeqn
which implies $\delta k \approx 32\,\mbox{MeV}/\hbar c$.
Nuclear excitations and shell-effects at the 1\,MeV energy implies
$\delta k \approx 4\,\mbox{MeV}/\hbar c$ and below.
From this perspective, the pion is a high-energy dof.
It will be important to trace how this argument plays out in diagrammatic constructions.

In the pseudopotential approach, coupling constants are fit to data (with constraints).
By deriving the EDF in Hartree-Fock form, self-interaction pathologies that plague beyond-mean-field calculations are avoided.
The resulting functionals pass tests for scale independence, convergence, and naturalness.
More recent pseudo\-pot\-en\-tials generate a
spuriousity-free nonlocal EDF that can describe pairing without den\-sity-de\-pendent terms~\cite{Bennaceur:2016hci}.
Finally, an approach to connect ab initio calculations to a pseudo-potential-based EDF
is laid out in ref.~\cite{Dobaczewski:2015eva}.
Here one derives the couplings not from experiment but from theoretical calculations of finite systems
using a constrained variation method, which is one way to define a density functional theory.

%%%%%%%%%%%%%%%%%%%%%%%%%%%%%%%%%%%%%%%%%%%%%%%%%%%%%%%%%%%%%%%%%%%%%%%%%%%%%%%%%%%%%

Finally, we consider the work by Saperstein and collaborators~\cite{Saperstein:2017hii},
which is rooted in Landau's Fermi liquid theory as extended to finite nuclei by Migdal~\cite{migdal1967theory,Migdal:1978az}.
The Fermi liquid approach has led to the very successful Fayans functionals, which
have significant phenomenological components~\cite{Bulgac:2017bho}.
But the underlying formalism is in the spirit of EFTs.
Indeed, Landau theory has been formalized by Polchinski~\cite{Polchinski:1992ed} and
Shankar~\cite{Shankar:1993pf} as an expansion about the Fermi surface.
If this can be extended to include the bulk properties, building on 
the \emph{self-consistent} theory of finite fermi systems in \cite{Saperstein:2017hii},
it would provide insight or even a direct path to a proper nuclear EDF.

%%%%%%%%%%%%%%%%%%%%%%%%%%%%%%%%%%%%%%%%%%%%%%%%%%%%%%%%%%%%%%%%%%%%%%%%%%%%%%%%%%%%%

\subsection{So what do we conclude?} \label{subsec:takeaways}

The many approaches touched on in this section give multiple avenues for making progress
toward improved EDFs.
I'm afraid it is not at all clear (to me) which should be favored and I do not propose to choose a winner.
But we have accumulated many clues about how we might choose dofs and formulate
an EFT for nuclear EDFs, so we will proceed with one particular path not yet followed.

%%%%%%%%%%%%%%%%%%%%%%%%%%%%%%%%%%%%%%%%%%%%%%%%%%%%%%%%%%%%%%%%%%%%%%%%%%%%%%%%%%%%%
%%%%%%%%%%%%%%%%%%%%%%%%%%%%%%%%%%%%%%%%%%%%%%%%%%%%%%%%%%%%%%%%%%%%%%%%%%%%%%%%%%%%%

\section{Effective action as a framework for proper EFT for EDFs} \label{sec:favorite}

Here I will lay out my own motivation and plan for making progress
on a proper EFT for nuclear EDFs, based on
a bottom-up EFT for DFT using an effective action formulation with auxiliary fields
and an explicit treatment of zero modes.

%%%%%%%%%%%%%%%%%%%%%%%%%%%%%%%%%%%%%%%%%%%%%%%%%%%%%%%%%%%%%%%%%%%%%%%%%%%%%%%%%%%%%

\subsection{Addressing questions about EFT for DFT}

Any path toward a proper effective theory for nuclear EDFs must confront some pertinent questions
about the general features of an EFT.   
Here are my prejudices at present for the answers based on insights from the explorations outlined in the previous section.
\begin{itemize}
   \item \emph{What are the optimal dofs?}
     The first point is whether the dofs of chiral EFT are appropriate.  My take is that the resolution
     of chiral EFT with nucleons and pions is too high for an efficient EFT, based on the smaller
     excitations scales at finite density and the success
     of the phenomenological EDFs. 
     A pionless theory expanded about free space does not naturally include the liquid drop systematics.
     But at finite density one could identify the dofs as quasinucleon densities 
     and the associated mean fields. 
     For low-lying collective degrees of freedom, time-dependent mean fields seem to be the
     natural choice. 
     A challenge to address is that quasiparticles are usually only considered as meaningful near the 
     Fermi surface, but we want a description for the bulk as well 
     (as suggested in \cite{Saperstein:2017hii}).
   \item \emph{Power counting: what is the expansion parameter for an EDF?}
     First, could it be the counting of chiral EFT?
     The fact that saturation in chiral EFT as currently formulated is driven by the repulsive three-body force, which does not
     appear until third order (N$^2$LO), suggests this is not a counting we want for heavy nuclei
     (so this reinforces an alternative
     choice of dofs).
     Indeed, we expect a proper EFT would produce saturation in leading order~\cite{Grasso:2016gls}.
     The work by Dyhdalo et al.\ on in-medium power counting with RG-softened interactions
     indicates how Pauli-blocking modifies the power counting~\cite{Dyhdalo:2017gyl}.
     Unfortunately, the actual expansion parameter for a phase-space-based approach
     is not obvious as yet.  When identified, 
     it should inform about the breakdown scale and therefore the range of applicability.
   \item \emph{How should the EDF as EFT be formulated?}
     One path is to use MBPT~\cite{Drut:2009ce}; I believe this is worth pursuing.
     However, as I will review in sect.~\ref{subsec:motivation}, effective actions are the natural and appropriate
      formalism for a field theoretic version of density functional theory (DFT)~\cite{FUKUDA94}. 
      This still leaves open multiple options for the implementation based on different
      types of effective action.
   \item \emph{How can we implement or expand about liquid drop physics?}
     The phenomenological EDFs are conventionally thought of as mean-field approximations, for which 
     the liquid drop physics and also shell structure are natural consequences.
     In an effective action formalism treated in a loop expansion, the leading piece is a
     saddlepoint contribution that defines mean fields and should improve with nuclear size as
     EDFs do. 
   \item \emph{What role should the renormalization group play?}  The renormalization group (RG) 
   should be an integral part of the ultimate EFT-for-EDFs formulation.
   For now we can highlight several areas where RG technology and insight is relevent.
   The first is the softening of interactions from decoupling high-momentum dofs.  As discussed
   in ref.~\cite{Drut:2009ce}, this is a natural prelude to nuclear EDFs that clearly affects
   the power counting of MBPT~\cite{Dyhdalo:2017gyl}. 
   The second is the use of RG to scale toward the Fermi surface
    in the EFT version of Landau theory~\cite{Polchinski:1992ed}; is something analogous 
    helpful for EDFs (\ie, a different type of scaling than with respect to the full momenta
    of nucleons in the medium)?
   The third is a top-down approach to \abinitio\ but orbit-free DFT 
    (i.e., not Kohn-Sham) proposed by Schwenk and Polonyi that uses a clever RG evolution~\cite{Schwenk:2004hm}. 
    The basic idea is to introduce an effective action for a nucleus with a low-momentum
    interaction included with a multiplicative factor $\lambda$ 
    and a confining background potential (\eg, a harmonic oscillator
    trap) with a factor $(1-\lambda)$.  As $\lambda$ flows from 0 to 1,
    the background potential is turned off and the interactions, with
    associated many-body correlations, are turned on.  This evolution
    is dictated by an RG equation in $\lambda$. 
    The first test implementations were by Braun and collaborators~\cite{Kemler:2013yka,Kemler:2016wci}
    and there are recent related developments by others~\cite{Liang:2017whg,Yokota:2018wue,PhysRevB.99.115106}.
    The practicality of this approach for nuclei will have to be demonstrated but at the least
    it should give valuable insight about
    the structure of a low-resolution functional and how to treat self-bound systems in DFT.

   \item \emph{How might we reconcile the different EDF representations via EFT?}
     One element is in the range of choices in implementing an EFT, such as the use
     of dimer fields in free-space pionless EFT as an alternative to only nucleon fields.  
     Thus we can use
     auxiliary fields or  expand in gradients with point couplings (contact terms).
     We also expect freedom from the choices how to regularize and renormalize, in analogy to the
     scale and scheme dependence of interactions in chiral EFT.
     Finally, there is also the change in representation possible from field redefinitions
     (or unitary transformations).

   \item \emph{How should we deal with symmetry breaking that is so integral to the standard EDF approach?} 
     In the context of effective actions, the symmetry breaking is precipitated by approximations
     stemming from a saddlepoint evaluation.  A consequence is zero modes when one expands about
     the saddlepoint --- part of what is called going ``beyond mean field'' in the EDF literature.  
     The zero modes must be dealt with as a priority for the EFT program to succeed; 
     my preferred plan is to adapt methods used for gauge theories
     to handle collective coordinates,
     as has been done in other contexts, although this is not yet shown to be a viable approach for DFT.
   \item \emph{Can we implement an EFT for EDFs without losing the favorable
            computational scaling of current nuclear EDFs?}  This is a pertinent question,
            but one that can't be answered at present.
            If the computational cost to evaluate the functional does increase significantly,
            there is the option to use statistical emulators (\eg, Gaussian processes).  
   \item \emph{What makes it possible to have a simple form for the phenomenological EFTs?}
      In Coulomb DFT this is related to the robustness of local density approximations (LDA).
      This realized in nuclear EDFs through a gradient expansion of densities.
      In field theoretic terms, the question becomes whether low-order operator product
      expansions can be used for the action.

\end{itemize}
It should be clear that these are provisional answers and may not be exclusionary, \eg, there may be more than
one reasonable choice of degrees of 
freedom.

%%%%%%%%%%%%%%%%%%%%%%%%%%%%%%%%%%%%%%%%%%%%%%%%%%%%%%%%%%%%%%%%%%%%%%%%%%%%%%%%%%%%%
\subsection{Schematic look at effective actions}

To implement a bottom-up EFT for EDFs, we need an appropriate formalism.
I believe the natural choice is an effective action formulation~\cite{FUKUDA94,Polonyi2001}.
Motivation will be given in sect.~\ref{subsec:motivation}, but first we
give a schematic overview of some of the key features, neglecting at first the
issue of zero modes.  
More detail can be found in ref.~\cite{Drut:2009ce} and references cited there.

If we focus only on ground states, we can build on the intuition physicists have
for ordinary thermodynamics with $N$ particles as
temperature $T \rightarrow 0$.
The thermodynamic potential is derived from the grand canonical partition
function, with the chemical potential $\mu$ acting as a  
source to change $N = \langle \wh N\rangle$,
 \beqn
    \Omega(\mu) = -kT \ln Z(\mu)
    \quad
    \mbox{and}
    \quad
    N = -\left(\frac{\partial\Omega}{\partial\mu}
   \right)_{TV} \; .
 \eeqn
Because $\Omega$ is convex, $N$ is a monotonically increasing
function of $\mu$ and we can
\emph{invert} to find $\mu(N)$ and apply a Legendre transform to obtain
 \beqn
   F(N) = \Omega(\mu(N)) + \mu(N) N \; .
 \eeqn
This is our (free) energy function of the particle number, which is 
analogous to the DFT energy functional of the density.%
\footnote{Because $\Vext$ is typically given rather than eliminated, for
a closer analogy we would also define $\Omega_\mu(N) \equiv F(N) - \mu N$,
which depends explicitly on both $N$ and $\mu$.  This gives the 
grand potential when minimized with respect to
$N$~\cite{Argaman:2000xx}.}

If we
generalize to a spatially dependent chemical potential $J(\xvec)$, then
\begin{align}
    Z(\mu) &\longrightarrow Z[J(\xvec)] \\
    \mu N = \mu\int\psi^\dagger\psi
      &\longrightarrow
      \int J(\xvec)\psi^\dagger\psi(\xvec) \; .
\end{align}
Continuing the analogy, 
we can do a functional Legendre transform from $\ln Z[J(\xvec)]$ to 
$\Gamma[\rho(\xvec)]$, where $\rho = \langle\psi^\dagger\psi\rangle_J$,
and we have DFT with $\Gamma$ simply proportional to 
the energy functional. 

The functional $\Gamma$ is one type of effective action~\cite{Polonyi2001}.
An effective action is generically the
Legendre transform of a generating functional
with an external source (or sources).  For DFT, we use a source to adjust
the density (\cf\ using an external applied magnetic field to
adjust the magnetization in a spin system).
Consider first the simplest case of a single 
external source $J(\bfx)$ coupled to the density operator 
$\wh \rho(x) \equiv \psi^\dagger(x)\psi(x)$ in the partition
function
\begin{align}
    \mathcal{Z}[J] = 
    e^{-W[J]} &\sim {\rm Tr\,} 
      e^{-\beta (\wh H + J\,\wh \rho) } \\
    &\sim \int\!\mathcal{D}[\psi^\dagger]\mathcal{D}[\psi]
    \,e^{-\int\! [\mathcal{L} + J\,\psi^\dagger\psi]} 
    \;,
    \label{eq:ZofJ}
\end{align}
for which we can construct a (Euclidean) path integral representation
with Lagrangian $\mathcal{L}$ \cite{Negele:1988vy}.
(Note: because our treatment is schematic, for convenience
we neglect normalization factors and take the inverse temperature
$\beta$ and the volume $\Omega$ equal to unity.)
The static density $\rho(\bfx)$ in the presence of $J(\bfx)$ is
\beqn
  \rho(\bfx) \equiv \langle \wh \rho(\bfx) \rangle_{J}
   = \frac{\delta W[J]}{\delta J(\bfx)}
   \;,
\eeqn  
which we invert to find $J[\rho]$ and then Legendre transform from $J$ to
$\rho$:
\beqn
   \Gamma[\rho] = - W[J] + \int\!d{\bfx}\, J(\bfx) \rho(\bfx) \;,
   \label{eq:gammarho}
\eeqn
with
\beqn
   J(\bfx) = \frac{\delta \Gamma[\rho]}{\delta \rho(\bfx)}
   \longrightarrow 
   \left.
   \frac{\delta \Gamma[\rho]}{\delta \rho(\bfx)}\right|_{\rho_{\rm gs}(\bfx)
   } =0
   \;.
   \label{eq:Jofx}
\eeqn 
For static $\rho(\bfx)$, $\Gamma[\rho]$ is proportional to 
the conventional Hohenberg-Kohn energy functional, which
by eq.~(\ref{eq:Jofx}) is extremized at the ground state density
$\rho_{\rm gs}(\bfx)$ (note that thermodynamic arguments establish that it is
a minimum \cite{Valiev:1997bb}).%
\footnote{A Minkowski-space formulation of
the effective action with time-dependent sources
leads naturally to an RPA-like generalization
of DFT that can be used to calculate properties of collective excitations.}

Consider the partition function in the zero-tem\-per\-a\-ture limit of
a Hamiltonian with time-independent source $J({\bf x})$
\cite{Zinnjustin:2002}:
 \beqn
   { \Hhat}(J) = {\Hhat}  + \int\!  J\, \psi^\dagger\psi \; .
   \label{eq:HhatJ}
 \eeqn
\emph{If} the ground state is isolated (and bounded from below),
 \beqn
   e^{-\beta \Hhat} = e^{-\beta E_0}
     \left[
       | 0 \rangle \langle 0 |
  + {\cal O}\bigl(e^{-\beta (E_1 - E_0)}\bigr)
     \right]
     \; .
     \label{eq:isolated}
 \eeqn
 As  $\beta \rightarrow \infty$, ${\cal Z}[J]$ yields the
ground state of ${\Hhat}(J)$ with energy $E_0(J)$:
  \beqn    
  { E_0(J)} = \lim_{\beta\rightarrow \infty} -\frac{1}{\beta} \log 
  {\cal Z}[J]
    = \frac{1}{\beta}W[J] \; .
 \eeqn
Substitute and separate out the pieces:
 \begin{align}
 E_0(J) &= \langle {\Hhat }(J) \rangle_J 
      = \langle \Hhat \rangle_J 
   + \int\! J \langle \psi^\dagger\psi \rangle_J \\
      &= \langle {\Hhat} \rangle_J + \int\! J\, \rho(J)
      \; .
 \end{align}
Rearranging,
the expectation value of ${\Hhat}$ in the ground state
generated by $J[\rho]$ is%
\footnote{The    
functionals will change with resolution or field redefinitions; 
only stationary points are observables.
This can be seen from eq.~(\ref{eq:expH}), where $\Gamma[\rho]$
is not the expectation value of $\Hhat$ in an eigenstate 
unless $J = J[\rho_{\rm gs}]$.}
 \beqn
   { \langle {\Hhat} \rangle_J} =  E_0(J) - \int J\, \rho
    = \frac{1}{\beta}\Gamma[\rho]
    \; .
      \label{eq:expH}
 \eeqn
Now put it all together:
   \begin{align}
      \frac{1}{\beta}\Gamma[\rho] &= \langle {\Hhat} \rangle_J 
     \stackrel{J\rightarrow 0}{\longrightarrow}
       E_0 \\
    J(x) &= -\frac{\delta \Gamma[\rho]}{\delta \rho(x)}
     \stackrel{J\rightarrow 0}{\longrightarrow}
    {
    \left.\frac{\delta \Gamma[\rho]}{\delta \rho(x)}
            \right|_{\rho_{\rm gs}(\bfx)} =0 }
      \; .
   \end{align}
So for static $\rho(\bfx)$, $\Gamma[\rho]$ is indeed proportional to 
the DFT energy density functional.  
Furthermore, the true ground state (with $J=0$) is a  variational
minimum,%
\footnote{For the Minkowski-space version of this discussion,
    see ref.~\cite{Weinberg:1996II}.}
so additional sources should be better than just
one source coupled to the density (these sources will couple to additional
densities such as the kinetic energy density in a Skyrme EDF).
The universal dependence on a non-zero external potential $v$ follows
directly in this formalism:
    \begin{align}
      \Gamma_{v}[\rho] &= W_{v}[J] - \int\! J\,\rho \\
       &= W_{v=0}[J+v] - \int\! [(J+v)-v]\, \rho \\
       &= \Gamma_{v=0}[\rho] + \int\! v\,\rho
       \; .
    \end{align}
Thus allowing for non-zero $\Vext$ is a trivial modification
to $\Gamma[\rho]$.  For the nuclear application, however, it is the
\emph{lack} of an external potential that complicates the formulation.  

To summarize, conventional microscopic DFT follows naturally
from calculating the response of a many-body system
to external sources, as in Green's function methods, only with local, static
sources that couple to densities rather than fundamental fields.  
(Time-dependent sources can be used for certain excited states.)
We can consider the zero temperature limit
of the partition function $\mathcal{Z}$ for
the (finite) system of interest in
the presence of external sources coupled to various quantities of
interest (such as the fermion density).
We derive energy functionals of these quantities by Legendre
transformations with respect to the sources \cite{Kutzelnigg:2006aa}. 
These sources probe, in a
variational sense, configurations near the ground state. 

The work by Lieb~\cite{Lieb:1983} on the Hohenberg-Kohn 
theorem~\cite{Hohenberg:1964zz} establishes that the critical issue 
for DFT is
the existence of the Legendre transform $F[\rho]$ of the
ground state energy as a functional $E[v]$ of the potential.
The details involve sophisticated
mathematics (\eg, convex-functional analysis) 
that is not readily accessible;
I  recommend ref.~\cite{Kutzelnigg:2006aa} by Kutzelnigg
as a gateway to the mathematically rigorous literature behind
DFT in terms of Legendre transformations.%
\footnote{There are important formal details~\cite{Eschrig:2003}, 
such as that we need $E[v]$ to
be concave to carry out the transform.}   

However, this attractive picture of DFT in field theoretic terms is not sufficient for the
nuclear case, because nuclei are self-bound systems.
The problems are clear even in our schematic treatment because we need the
source to couple only to internal dofs.
For now we will assume this will all work out and return to discuss a solution
in sect.~\ref{subsec:zeromodes}.

%%%%%%%%%%%%%%%%%%%%%%%%%%%%%%%%%%%%%%%%%%%%%%%%%%%%%%%%%%%%%%%%%%%%%%%%%%%%%%%%%%%%%

%%%%%%%%%%%%%%%%%%%%%%%%%%%%%%%%%%%%%%%%%%%%%%%%%%%%%%%%%%%%%%%%%%%%%%%%%%%%%%%%%%%%%

\subsection{Motivation for effective action EFT for EDFs} \label{subsec:motivation}

Here I enumerate the advantages of the effective actions in path integral form for implementing
an EFT for nuclear EDFs~\cite{Drut:2009ce}.  Note that because there will be an underlying
EFT Hamiltonian, these EDFs will be generalizations of the functionals of DFT.

  \begin{itemize}
  \item Effective actions are the natural theoretical framework for Legendre
  transforms~\cite{Peskin:1995ev,Weinberg:1996II,Zinnjustin:2002}, 
  which is the underlying basis for DFT.  Note that these aspects
  tend to be hidden when using wave function methods.
  \item The path integral construction of DFT is transparent, such as
  the role and usefulness of additional densities/sources.  
  \item Path integral effective actions are particularly 
  suited for symmetry breaking, such as encountered with
  pairing.  The renormalization issues in pairing are manifest rather than hidden.
  \item Connections to the RG, and therefore to EFT and power
  counting, can be more accessible.
  \item The path integral formulation puts the DFT construction
  in a broader perspective, which can suggest connections and
  generalizations not apparent otherwise.  For example, there
  are alternative effective actions using auxiliary fields or
  with a two-particle-irreducible nature.
  The former will be considered in sect.~\ref{subsec:auxiliary}.  
  The latter may be related to more general EDF constructions
  proposed in ref.~\cite{Duguet:2009gc}.
  \item
  The quantization of
  gauge theories was greatly facilitated by Faddeev-Popov 
  and BRST methods
  using path integrals; analogous techniques offer alternative
  possibilities for implementing collective coordinates to robustly
  address the issue of
   symmetries broken at the mean-field level~\cite{Bes:1990} 
  (see sect.~\ref{subsec:zeromodes}).
  \item The path integral formulation can suggest different
  types of nonperturbative approximations, such as $1/N$
  expansions, that can organize for extensions beyond the mean-field level.
\end{itemize}

In summary, effective actions are a natural framework to implement Legendre
transformations, motivate approximations not obvious in MBPT,
and allow us to consider generalizations of conventional DFT.
One limitation of DFT is the exclusive
role of local potentials (sources)
and densities, where locality is in reference to coordinate
space.
Kutzelnigg points out that this is in contrast to many-body
methods that introduce a finite basis in which operators are expanded,
for which local operators have no privileged place.  In this sense,
density matrix functional theory, as proposed for nuclei
in ref.~\cite{Duguet:2009gc}, seems more
natural~\cite{Kutzelnigg:2006aa}.  
By looking at effective actions as a broader context, the limitations
and problems of local sources are apparent, but also the opportunities for 
generalizations.

%%%%%%%%%%%%%%%%%%%%%%%%%%%%%%%%%%%%%%%%%%%%%%%%%%%%%%%%%%%%%%%%%%%%%%%%%%%%%%%%%%%%%
%%%%%%%%%%%%%%%%%%%%%%%%%%%%%%%%%%%%%%%%%%%%%%%%%%%%%%%%%%%%%%%%%%%%%%%%%%%%%%%%%%%%%

\subsection{Auxiliary fields} \label{subsec:auxiliary}

     I propose a bottom-up construction, building on the insights from the
phase-space approach,  the successful phenomenology of nuclear EDFs, and 
the DFT framework.
In particular, I would like to build in that the pairing and particle-hole channels seem to have
different associated scales and that collective modes are dominant low-energy dofs beyond the mean fields.
To accomplish both, we can look to textbook treatment for analogous condensed matter systems, which
introduce auxiliary bosonic fields for fermion bilinears via Hubbard-Stratonovich (HS) 
transformations~\cite{FUKUDA94,Nagaosa:1999,Altland:2006,Furnstahl:2002gt}.

To my knowledge, this path for nuclei has not yet been explored. 
In previous work with my collaborators, 
we studied effective actions of composite operators, because that closely paralleled the
form of Skyrme functions.
This required use of the so-called ``inversion method'' to carry out the Legendre transformation.
It seemed well adapted to EFT because the inversion was carried out order-by-order in an expansion.
Pairing was included by coupling sources not only to the proton and neutron densities but to
the anomalous density~\cite{Furnstahl:2006pa,Furnstahl:2007xm,Furnstahl:2008df}.  
This meant, however, that the particle-hole and pairing channels were treated
on the same footing.
It also meant that collective excitations would not be naturally included in a loop expansion
but would have to be built up from coherent nucleon
pairs at different orders.

In the auxiliary field (HS) method~\cite{Nagaosa:1999,Stone:2000}, 
one couples operators such as $\psi^\dagger\psi$ to an auxiliary field $\varphi$,
and eliminate all or part of $(\psi^\dagger\psi)^2$.
The Legendre transform follows from adding a source term $J\varphi$ and performing a 
loop expansion about the expectation value $\phi=\langle\varphi\rangle$.
The inversion is direct~\cite{Peskin:1995ev}, in contrast to the case of composite operators where
a perturbative inversion method is needed.
We integrate out the fermion fields, 
whence a saddlepoint expansion gives a mean-field approximation
in leading order 
(with freedom to choose how this is organized, \eg, Hartree, HF, HFB~\cite{Negele:1988vy})
and then the next order has collective contributions (corresponding to the RPA).
In practice for nuclei we evaluate the fermion
determinant using
solutions to Schr\"odinger equations for single-particle quasi-nucleon wave functions
in the mean fields; this
generates the Kohn-Sham system.
We can use the freedom of the expansion to require the density be unchanged at each order.

\begin{figure}[th!]
        \begin{center}
        \includegraphics[width=0.98\columnwidth]{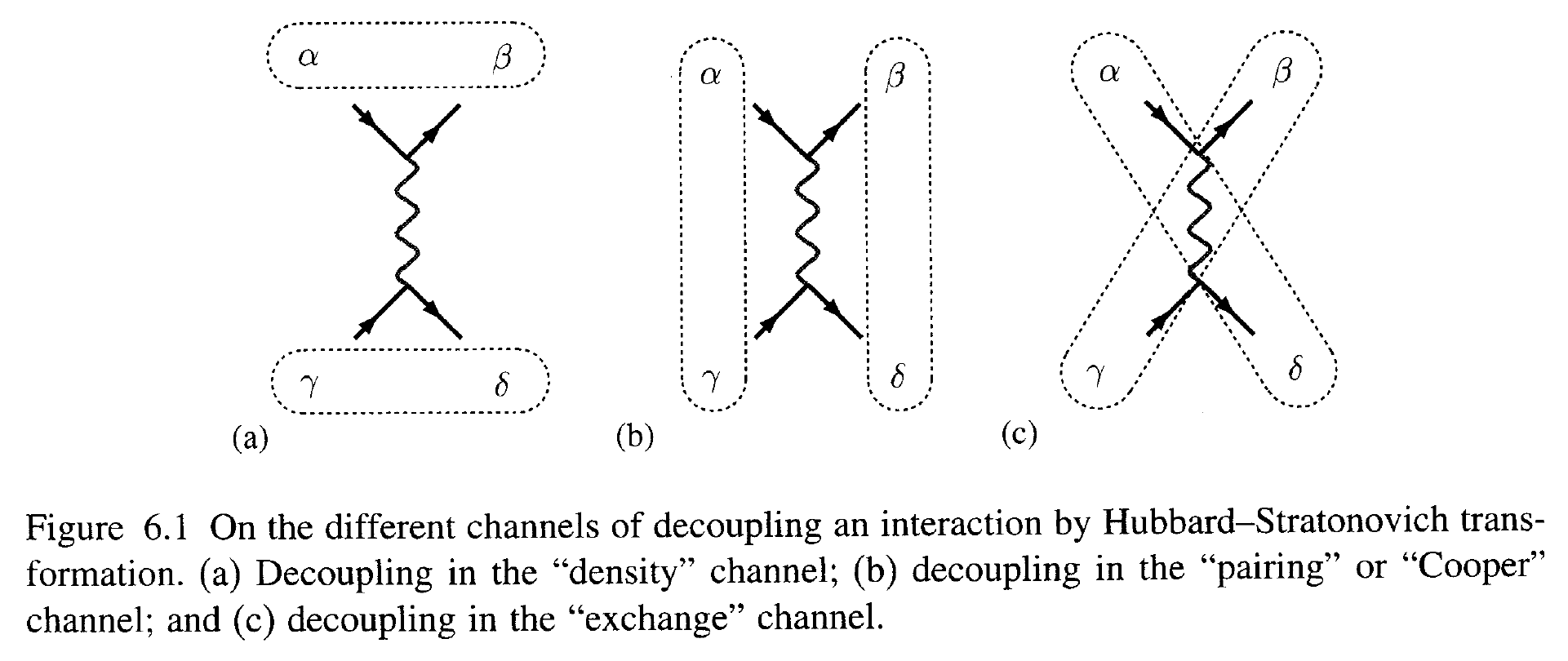}
        \caption{Decoupling in different channels~\cite{Altland:2006}.}
        \label{fig:decoupling}
        \end{center}
\end{figure}

In an EFT approach that restricts momenta, 
one would directly introduce a complete (but not redundant) set of bosonic operators
coupled to quadratic fermion bilinears.
The ``trick'' in this case is to introduce fields for all of the channels, not just one
(they are indicated schematically for two-body interactions in fig.~\ref{fig:decoupling}). 
If one
summed contributions from all momenta to all orders, this would lead to double counting.  
But for small momenta near the Fermi surface,
the different channels are effectively independent (these are different small momenta in the
different channels), so one can and should include both~\cite{Nagaosa:1999,Altland:2006}.
This would then be 
consistent with nuclear EDF phenomenology.

The first steps for testing whether this formulation is viable is to apply it to a 
uniform system (an effective action treatment with one auxiliary field was explored in \cite{Furnstahl:2002gt}) 
and then to extend to a system in a trap. 
It is possible that the subtle renormalization issues arising with pairing 
when coupling to composite densities~\cite{Furnstahl:2006pa} will be easier to deal with
in terms of the auxiliary fields.
We also note that the proposed approach is close to phenomenological covariant DFT phenomenology with meson fields.
The same sort of self-consistency cycle with be relevant: given mean fields, solve equations for 
the quasi-nucleon wave functions, solve the mean-field equations, and repeat until converged.
We also highlight the connection to Migdal's treatment of finite Fermi 
systems~\cite{migdal1967theory,Grummer:2006we}.

Of course, all of this will only be useful if we can carry out a tractable way to handle zero modes.

%%%%%%%%%%%%%%%%%%%%%%%%%%%%%%%%%%%%%%%%%%%%%%%%%%%%%%%%%%%%%%%%%%%%%%%%%%%%%%%%%%%%%

\subsection{Dealing with zero modes} \label{subsec:zeromodes}

Ordinary nuclei are self-bound, which presents conceptual issues
about whether Kohn-Sham DFT is well defined and practical problems
on how to deal with the consequences of symmetry breaking by the
Kohn-Sham potentials, which will not have 
all of the symmetries of the Hamiltonian~\cite{Blaizot:1985}.
These broken symmetries include the $U(1)$ phase symmetry for fermion
number and translational and rotational invariance.

In conventional discussions of EDFs the issue of symmetry breaking plays 
an important role; indeed a distinction between DFT and the EDF method is made on
this basis~\cite{Duguet:2010cv}.
The dilemma with properly treating symmetries in the nuclear many-body problem is that
one wants simple wave functions (\eg, Slater determinants) but this misses correlations
from symmetries (\eg, plane waves won't describe clustering into nuclei)~\cite{Ring:2005}.
In developing nuclear DFT or some 
variation of it within a field theory (effective action) formalism and as an EFT,
the symmetry breaking manifests as zero modes.
In particular these arise
when one does a saddlepoint or stationary phase expansion of the path integral for the nuclear ground state,
which in turn leads one to pick out a mean-field reference state. 
The quantum corrections will naively be
found to be infinite because there are fluctuations possible in flat (symmetry-wise) directions.
Mathematically, one must calculate a determinant when evaluating quadratic fluctuations and 
zero eigenvalues (hence ``zero modes'') will cause divergences.

The textbook discussions of how to restore mean-field broken
symmetries tend to follow one of these two related lines of discussion:
\bi
 \item
States connected by a unitary transformation $U(\alpha)$
corresponding to a broken symmetry are degenerate:
\beqn
   | \phi\,\alpha \ra = U(\alpha) | \phi \ra
\eeqn
with $|\phi\ra$ a  ``deformed'' state, implies
\beqn
  \la \phi\,\alpha | H_N | \phi\,\alpha\ra = \la \phi | H_N | \phi \ra 
  \;.
\eeqn
The degeneracy can be removed by diagonalizing in the subspace
spanned by the degenerate states.
The group parameter
$\alpha$ for continuous groups can be considered a
\emph{collective coordinate}, which specifies the
orientation in gauge space of the deformed state $|\phi \ra$.
A general strategy is to transform from $3A$ particle coordinates
into collective and internal coordinates~\cite{Ring:2005}. 
 
 \item
In finite systems, broken symmetries arise only as a result of 
approximations.  This usually happens with variational
calculations over trial wavefunctions that are too restricted;
a mean-field approximation is an example.
The symmetry can be restored by using a linear
superposition of degenerate states:
\beqn
  | \psi \ra = \int\! d\alpha\, f(\alpha) | \phi\,\alpha\ra
  \;,
\eeqn
which when minimized with respect to the $f(\alpha)$'s pro\-jects
states of good symmetry~\cite{Blaizot:1985}.  
(Because minimizing with respect to 
$|\phi\ra$ and with respect to $f(\alpha)$ do not commute, there
are two types of projection.  It is most accurate to project
first and then find the best deformed state corresponding
to a given quantum number.)
For example, particle number projection for EDF's
is described in refs.~\cite{Stoitsov:2006tv,Dobaczewski:2007ch}.
\ei
When implemented, these approaches  are considered to
be beyond EDF, where there
are only densities and not a wavefunction.
From a different perspective, the restoration of broken symmetries
of GCM-type configuration mixings can be considered as
a ``multi-reference'' extension of the usual ``single-reference''
EDF implementation 
(see refs.~\cite{Lacroix:2008rj,Bender:2008rn,Duguet:2008rr}).

For nuclear DFT, the conceptual question was highlighted 
by Engel~\cite{Engel:2006qu}, who pointed out that the
ground state of a self-bound system, with a plane wave describing
the center of mass, has a density distributed uniformly over
space.  Clearly this is not the density one wants to find from DFT, 
so there is a question of how to proceed.
There are two separate considerations:  i) Does Kohn-Sham
DFT exist in a useful form for self-bound systems? 
ii) If so, how does one formulate and implement it?   
Engel and other authors have addressed this 
issue some time ago~\cite{Engel:2006qu,Giraud:2007pe,Barnea:2007jx,Giraud:2008zz,Giraud:2008yw,Messud:2009jh,Chamel:2010ac,Duguet:2010cv,Messud:2012bx},
with a consensus that HK existence proofs for DFT are still well
founded, but for \emph{internal} densities (\eg, meaning independent of the
center-of-mass motion when considering broken translational symmetry).%
\footnote{In other contexts, such densities are called ``intrinsic'',
but this has a different meaning in the context of symmetry breaking,
so ``internal'' is typically used instead.}

Wave function methods have
several related strategies for dealing with
the ``center of mass'' (COM) problem:
  \begin{enumerate}
    \I Isolate the ``internal'' degrees of freedom, typically by
    introducing Jacobi coordinates.  Then the observables are by
    construction independent of the COM.
    This gets increasingly
    cumbersome with greater numbers of particles.
    \I Work in a harmonic-oscillator Slater determinant basis,
    for which the COM decouples,
    and introduce a potential for the COM that allows its contribution
    to be subtracted.
    \I Work with the internal Hamiltonian (\ie,
    subtract the COM kinetic energy $T_{\rm CM}$)
    so that the COM part factorizes and does not contribute to
    observables to good approximation (see
    in particular ref.~\cite{Hagen:2009pq} for coupled
    cluster calculations). 
  \end{enumerate}
Versions of the first two possibilities are in fact
among the ideas considered for DFT in
refs.~\cite{Engel:2006qu,Giraud:2007pe,Barnea:2007jx,Giraud:2008zz,Giraud:2008yw,Messud:2009jh}.

For the effective action approach, 
the issue of broken symmetry was first addressed long ago in
the study of solitons~\cite{Rajaraman:1982,Negele:1988vy}, where it also
arises as the problem of dealing with zero modes when calculating quantum
fluctuations.  Methods found in the literature include those similar
to the textbook treatments of mean-field broken symmetry cited above.
A subsequent field-theoretic functional approach used a Fadeev-Popov construction to
introduce collective coordinates with ghost degrees of freedom (recall that
these ghost fields are spin-zero but anticommuting)~\cite{Rajaraman:1982}.
This works, but is cumbersome, particularly if the symmetry is non-Abelian.
For quantizing gauge theories, a more effective approach uses BRST symmetry.

BRST can be applied to the collective coordinate methods by building on the
equivalence of a theory with constraints and a gauge theory.  
It is then just another way to deal with the original constrained system, 
by imposing an eigenvalue condition on the states of the theory
to eliminate spurious dofs~\cite{Nemeschansky:1987xb}.
There exists a substantial literature, primarily from Bes and collaborators
(see ref.~\cite{Bes:1990,Bes:2016hjm} and references therein), on how to 
apply the BRST methods to translational, rotational, and pairing dofs.
They have demonstrated both variational and perturbative
approximations, mostly using algebraic methods.

The procedure starts with adding
fields for the collective degrees of freedom and Lagrange multipliers, so there is an overcomplete 
Hilbert space.
But rather than project out physical dofs, one doubles down and adds still more fields:
ghost fields and their conjugate momenta.
This might seem to complicate things, but in fact it simplifies them because there is
a supersymmetry between field variables and ghosts, called BRST symmetry~\cite{Peskin:1995ev}.
As in other contexts, the ghost fields serve to cancel spurious contributions, as enforced
by the BRST symmetry.
Associated with the symmetry is a charge $\Qhat$.

For a gauge theory, one can ensure physical results
by working with gauge-invariant operators in the Lagrangian or
Hamiltonian.  With the BRST, the focus is instead on
BRST-invariant operators, \ie, to exploit BRST symmetry rather than the original gauge
invariance.  
The trick of the BRST technique is to replace
the notion of a gauge transformation that shifts operators by c-number functions with
a BRST transformation that mixes operators with different statistics.  Gauge fixing is accomplished
by adding a BRST-invariant function to the Hamiltonian that is not invariant under an ordinary gauge
transformation and then work with $\HBRSThat$.
Physical states have $Q=0$ and
matrix elements of $\HBRSThat$ in these states will give the physical internal energy.

What this means for the broken-symmetry application is that $\HBRSThat$
commutes with the generator of collective transformations, so it displays unbroken collective symmetry,
and states are labeled by the corresponding quantum numbers.
But $\HBRSThat$ does not commute with the symmetry generator because of the gauge-fixing function (they
differ by a so-called null function), so there are no zero modes, 
no infrared divergences.
In this way there is a unique ground state and projection is automatically achieved.
The BRST approach has commonality with the features sought for nuclear DFT in refs.~\cite{Engel:2006qu,Barnea:2007jx,Messud:2009jh}, such as maintaining  the full set of orbitals.

Furthermore, it is ok if due to approximations the state is not exactly a $Q=0$ state, because approximately broken supersymmetry
does not lead to zero modes.  
So in practice states will be strictly states of the collective generator but approximate eigenstates
of $\Qhat$.  This doesn't disturb the favorable features and a better approximation means better
projection.  
Of course, it remains to be seen if suitable approximations are going to be feasible.

We can now imagine how eq.~(\ref{eq:HhatJ}) would be modified in the BRST framework.  
By replacing $\Hhat$ by
$\HBRSThat$ and then
coupling external sources to a BRST-invariant
field combination to probe near the ground state, one stays within the sector of Hilbert
space with internal (physical) dofs.
Thus the conditions to carry out the Legendre transformation and have a variational
state can be fulfilled.
The details have not been worked out before for this case
and there are both subtleties and unexplored freedom, so there is still much to be done.

%%%%%%%%%%%%%%%%%%%%%%%%%%%%%%%%%%%%%%%%%%%%%%%%%%%%%%%%%%%%%%%%%%%%%%%%%%%%%%%%%%%%%
%%%%%%%%%%%%%%%%%%%%%%%%%%%%%%%%%%%%%%%%%%%%%%%%%%%%%%%%%%%%%%%%%%%%%%%%%%%%%%%%%%%%%

\section{Outlook}  \label{sec:summary}

This is an exciting time in low-energy nuclear theory, with steadily improving calculations of structure
and reactions.  
There are multiple opportunities for improving the performance energy of the density functional
method;
I believe the best overall strategy at this stage is to pursue all promising avenues.
People are focusing on the relevant questions and developing impressive technology
that opens doors to new approaches.

The progress toward formulating a proper effective field theory to supplant nuclear EDFs 
has been less obvious.  I have proposed a path to follow toward an effective action
formulation, but now is just the
beginning of the journey.
As indicated, the most critical step in this program is implementing the BRST approach.
There have been several applications to simple quantum mechanical models to illustrate how
it works, although in an operator formalism.
There are also illuminating toy models for path-integral effective actions (even as
zero-dimensional field theories).  
The plan is to start by merging the path-integral toys with the toy models in refs.~\cite{Bes:2002ajp,Nemeschansky:1987xb} for proofs-of-principle and then 
graduate to simple models like those considered in refs.~\cite{Engel:2006qu,Bertolli:2008uq} 
before finally considering realistic nuclei.

\begin{acknowledgement}

I acknowledge many illuminating discussions over many years with my colleagues on effective field theory
and energy density functionals that have contributed to my reflections here.
However, all misunderstandings, misstatements, and misinterpretations are my own.
Supported in part by the US National Science Foundation
under Grant No.~PHY--1614460 and the NUCLEI SciDAC Collaboration under
US Department of Energy MSU subcontract RC107839-OSU\@.

\end{acknowledgement}

% \begin{table}
% \caption{Please write your table caption here}
% \label{tab:1}       % Give a unique label

% % For LaTeX tables use
% \begin{tabular}{lll}
% \hline\noalign{\smallskip}
% first & second & third  \\
% \noalign{\smallskip}\hline\noalign{\smallskip}
% number & number & number \\
% number & number & number \\
% \noalign{\smallskip}\hline
% \end{tabular}
% % Or use
% \vspace*{5cm}  % with the correct table height
% \end{table}
%

% BibTeX users please use
%\bibliographystyle{epj}
%\bibliography{dft_refs_old_review}
%

\end{document}